# Highlights from RHIC
# Plus some earlier BNL Highlights in Subnuclear Physics

Michael J. Tannenbaum*
Brookhaven National Laboratory
Upton, NY 11973 USA

## A new periodic table for the 21st Century

In the 20th century, the periodic table, which was invented in the 19th century, was understood using quantum mechanics, atomic, nuclear and subatomic physics; and many elements were added (Figure 1a). Also in the 20th century, the discovery of a whole new subnuclear world of physics totally changed our view of nature and led to a new periodic table to be understood (Figure 1b). The neutrons and protons, of which atomic nuclei

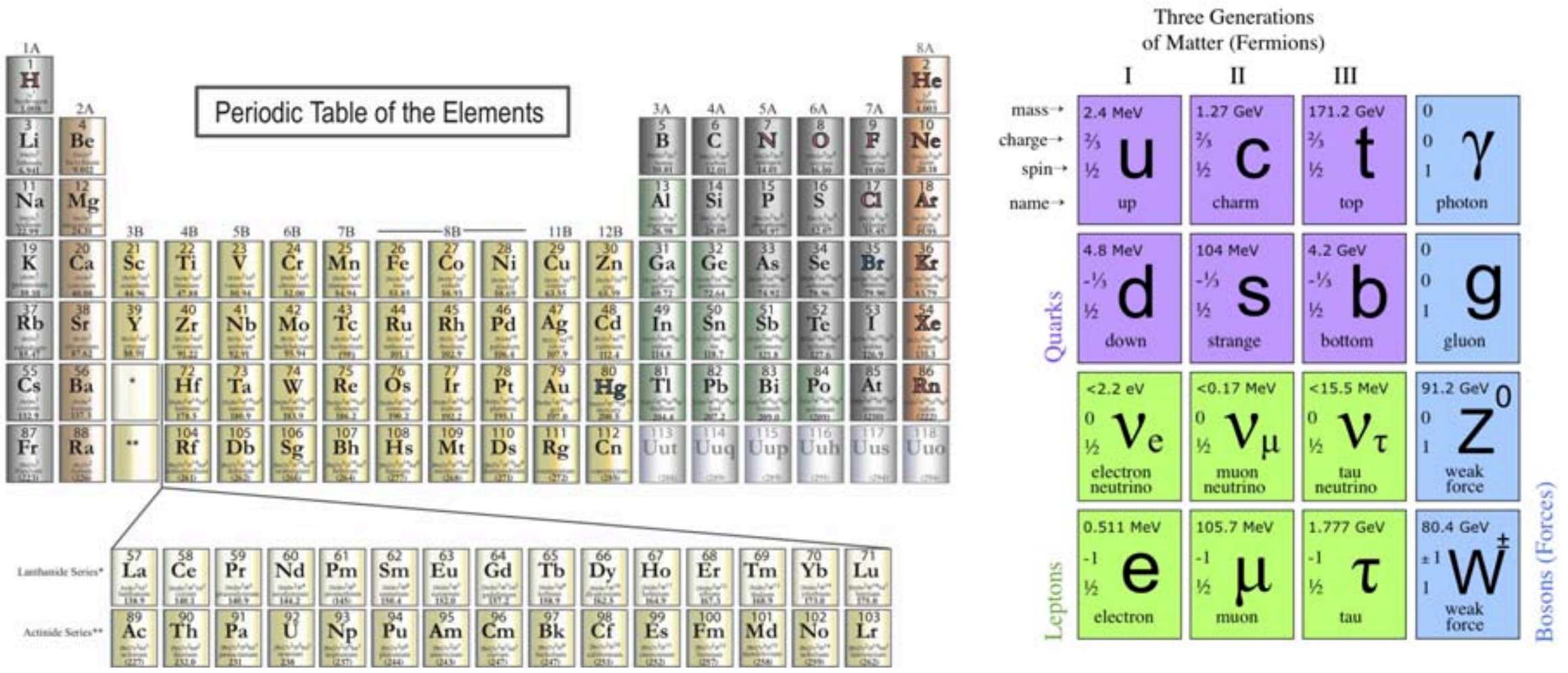

Figure 1. a) (left) Periodic Table of the Chemical Elements end of 20th Century (LANL 2011). b) (right) 21st century periodic table of the Quarks and Leptons (Fermions) and force carriers (Bosons) (FNAL 2011)

are composed, are themselves composed of the subconstituent up (*u*) and down (*d*) quarks which are held together by gluons, the quanta of the strong interaction in a theory called Quantum Chromodynamics or QCD (QCD 1983).

QCD has three charges called colors, e.g. red, green, blue; and in sharp distinction to elctromagnetism, whose quantum, the photon, carries no electric charge, the gluons of QCD are charged and therefore couple to each other. This provides the confinement of quarks and gluons inside nucleons as well as as a Coulomb-like $1/r^2$ short range force which can be measured. Each nucleon is composed of 3 quarks of different colors (*uud* for the proton and *udd* for the neutrons) so that the nucleons are color neutral. The nuclear force which binds neutrons and protons into nuclei is produced by a Yukawa coupling (Yukawa 1935) due to the exchange of pions which are made of quark-anti quark pairs, e.g $\pi^+$ ($u\bar{d}$).

* Supported by the U.S. Department of Energy, Contract No. DE-AC02-98CH1-886.

# Brookhaven National Laboratory (BNL)

I start with a brief description of Brookhaven National Laboratory and its contributions to the new periodic table, which concern the charm and strange quarks, the muon neutrino, and the electron neutrino. Brookhaven National Laboratory (BNL), a National Laboratory funded by the U.S. Federal Government, was founded by nine major northeastern universities in 1947 to promote basic research in the physical, chemical, biological and engineering aspects of the atomic sciences and for the purpose of the design, construction and operation of large scientific machines that individual institutions could not afford to develop on their own. It is located on Long Island roughly 100km east of New York City. In fact the Relativistic Heavy Ion Collider (RHIC) at BNL can be seen from outer space since it is not buried in a tunnel but is in an enclosure on the surface, which is covered by earth for shielding (Fig. 2).

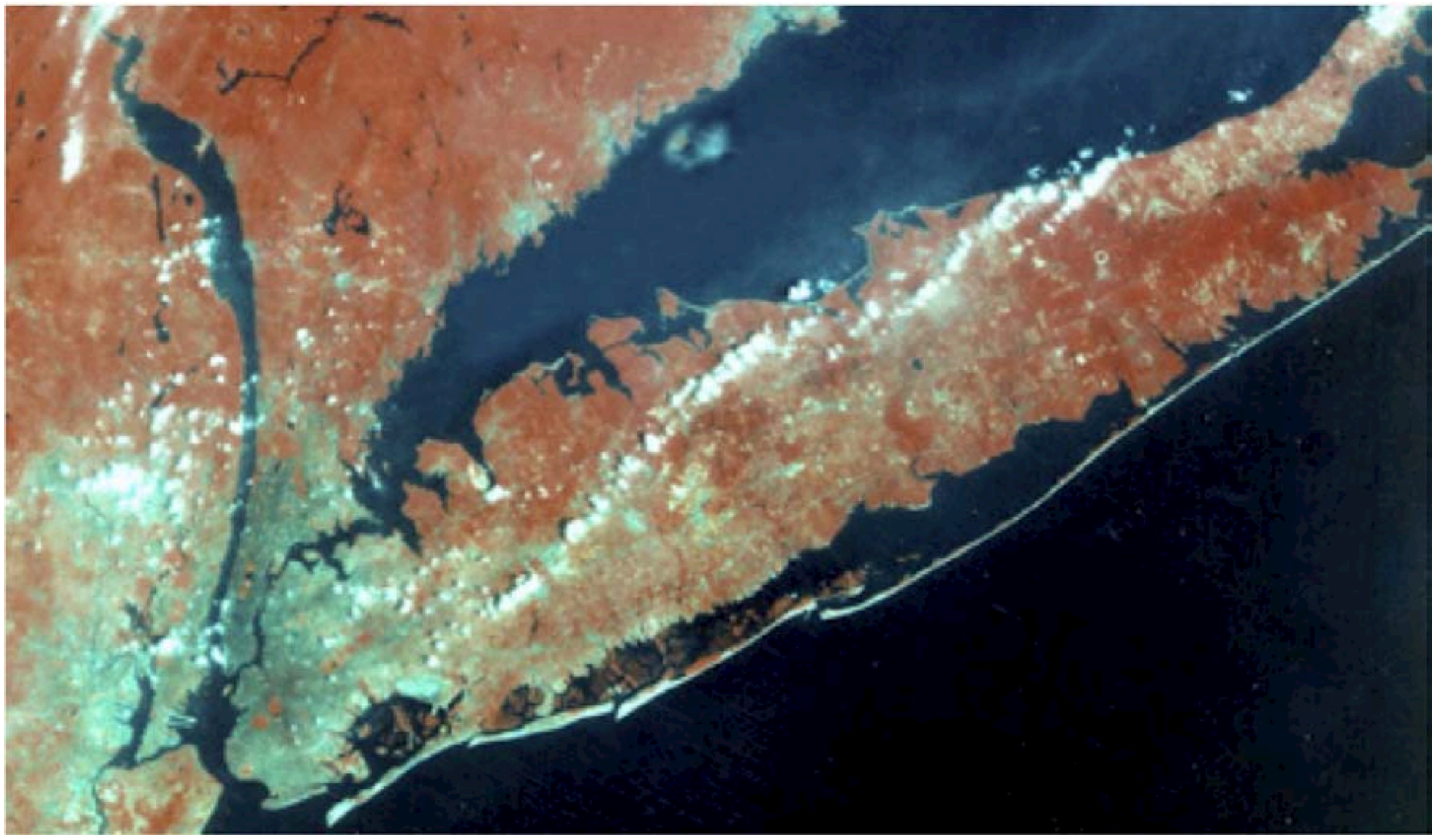

Figure 2. NASA Infra-red photo of New York Metro Region. RHIC is the white circle in the center of Long Island below the rightmost group of clouds. Manhattan island is clearly visible on the left side.

A closer aerial view of BNL (Fig. 3) shows the Relativistic Heavy Ion Collider (RHIC), which started operations in the year 2000, as well as the two previous accelerators built for High Energy Physics (HEP): the Cosmotron, a 3.3 GeV proton accelerator, which operated from 1953—1966, and the Alternating Gradient Synchrotron (AGS), which started operation in 1960 and ended operations for HEP in 2002 but now serves as the injector to RHIC.

Major discoveries and original contributions to the new periodic table were made at BNL during both the Cosmotron and AGS era. I shall briefly review these contributions before moving on to the highlights from RHIC.

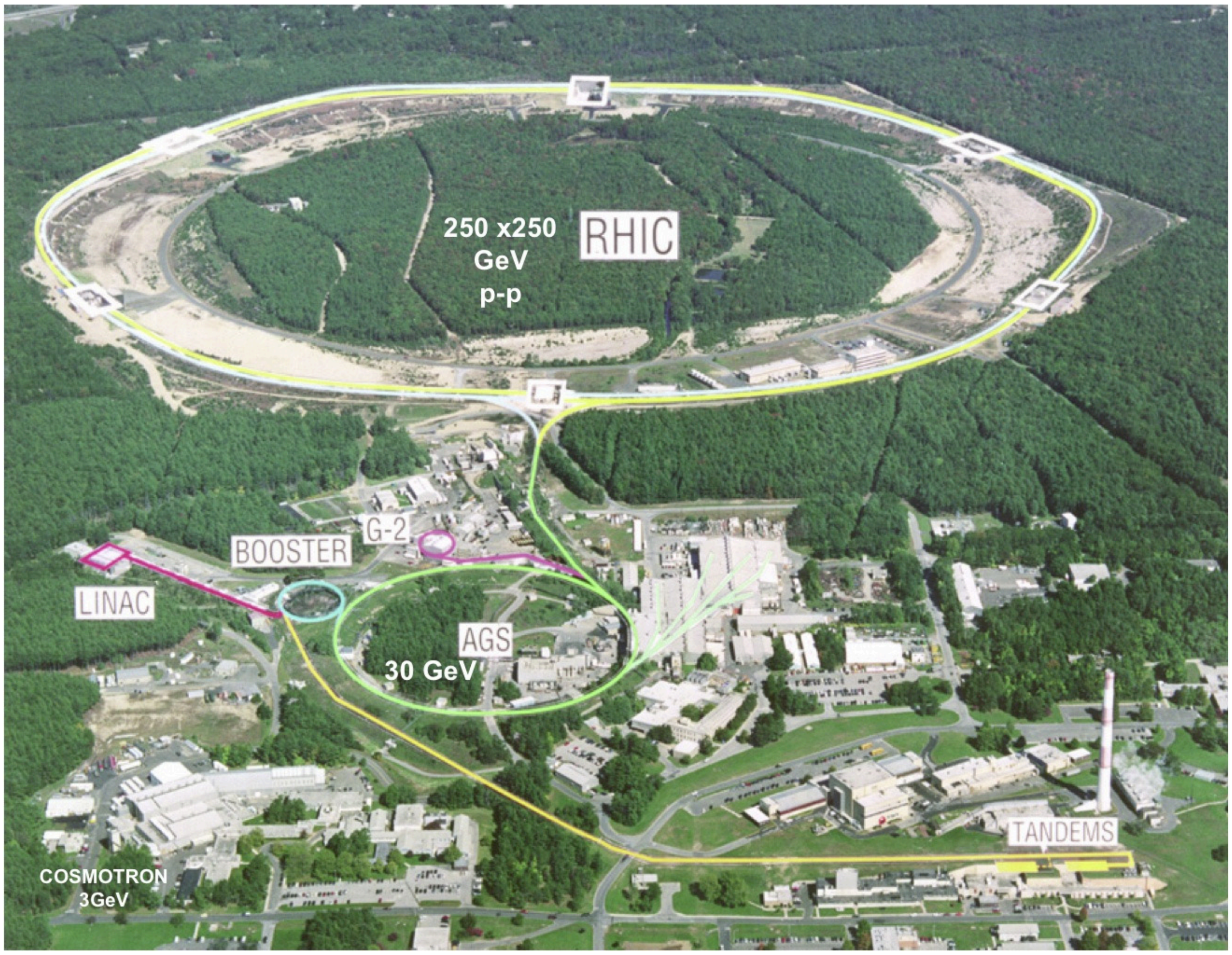

Figure 3. A closer view of RHIC at Brookhaven National Laboratory.The large circle without tree cover is excavation related to the enclosure of the RHIC machine. The labeled lines show the Linac, Booster accelerator for polarized proton injection, the tandem van de graaf accelerator and transfer line to the booster, and the AGS which accelerates the beams to an energy of 22 GeV per nucleon × Z/A where Z and A are the atomic number and weight of the nucleus.

## The Cosmotron Era 1953—1966

The Cosmotron (Figs. 3,4) was the first particle accelerator to deliver beams with energy greater than 1 Billion Electron-Volts (1 GeV), with a maximum accelelerated proton energy of 3.3 GeV. It was the first synchrotron with an extracted beam and the first accelerator to produce in the laboratory all the types of particles known from cosmic rays, including the strange particles, which were called strange because they were produced at a large rate, consistent with strong interactions, but decayed slowly, consistent with weak interactions. These strange particles were the first evidence for the strange (s) quark.

Some highlights of discoveries at the Cosmotron are:

- Associated production of strange particles, $\pi^-+p\rightarrow\Lambda^0 + K^0$ (Fowler 1954)[1]
- Multiparticle production in a n+p collision (Fowler-np1954)
- Long-lived, CP-odd, $K^0_2$. Limit $K^0_2\rightarrow\pi+\pi-$ <1% (Lande 1956)
- $\rho$ vector meson (Erwin 1961)

1. According to Pais (1986), “As to associated production, cosmic ray evidence seemed at first against it. ...The issue was settled when accelerators in the GeV range became available. A Cosmotron experiment [(Fowler 1954)] yielded the first convincing results.”

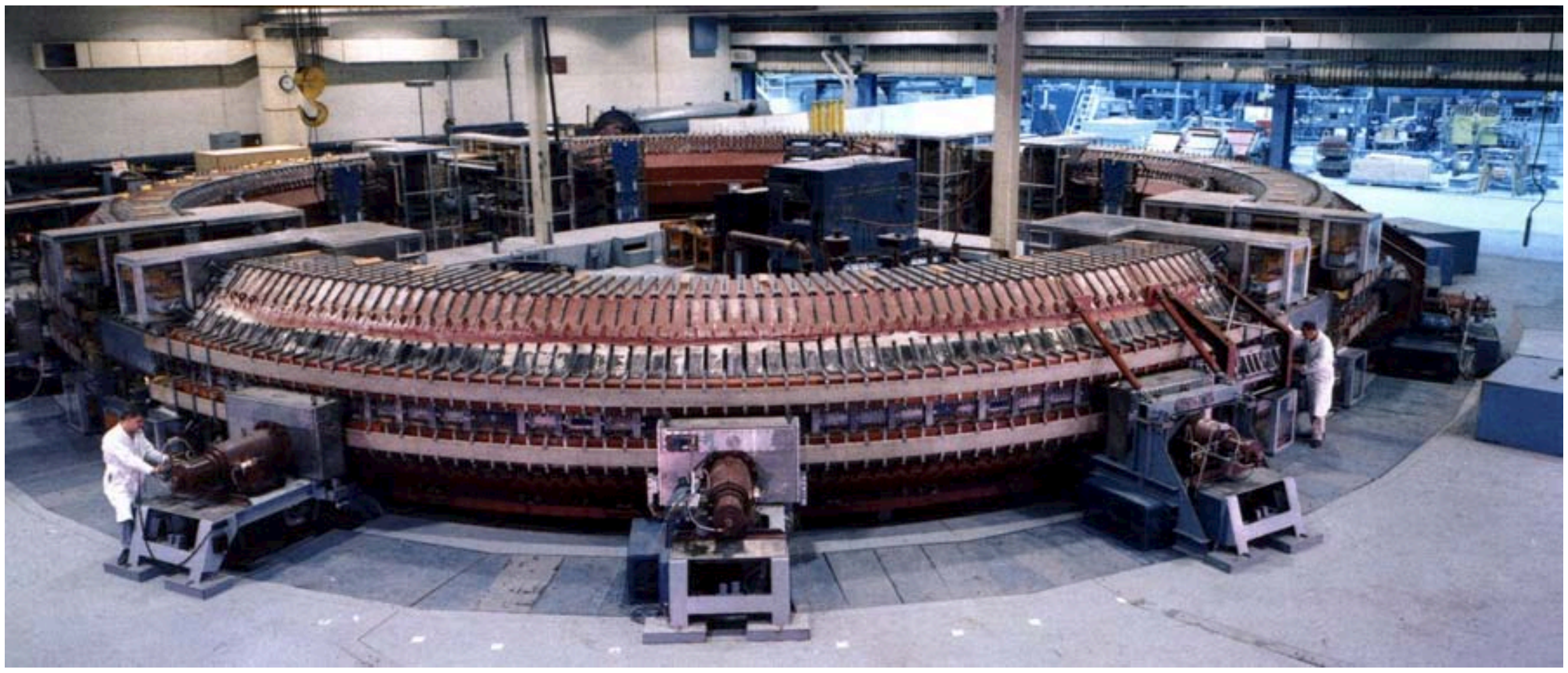

Figure 4. The Cosmotron machine. The machine is 75 feet in diameter, composed of 288 C-shaped magnets with the open gaps all facing outwards. The edge of the magnet coils are visible above and below the vacuum tank in the mid-plane, which is connected to the vacuum pumps. The machine is small enough to fit in a standard high-bay building. The machine was shielded with concrete blocks during operation.

There were also two major theory papers by BNL authors during this period:

- Gauge Field Theory. This is the basis of the standard model. (Yang-Mills 1954)
- Parity Violation in Weak Interactions? (Lee-Yang 1956) Nobel Prize 1957.

## Strong Focusing, discovered in 1952, leads to BNL-AGS (and CERN-PS).

During the last year of Cosmotron construction, a new principle of accelerator design was discovered at BNL (Courant 1952). Instead of C-magnets with their gaps all facing outwards, the orientation was alternated so that groups of magnets had gaps facing outwards and adjacent groups had gaps facing inwards. This provided, “A sequence of alternately converging and diverging magnetic lenses of equal strength [which] is itself converging, and leads to significant reductions in oscillation amplitude, both for radial and axial displacements, i.e. much smaller magnets.” (Courant 1952).

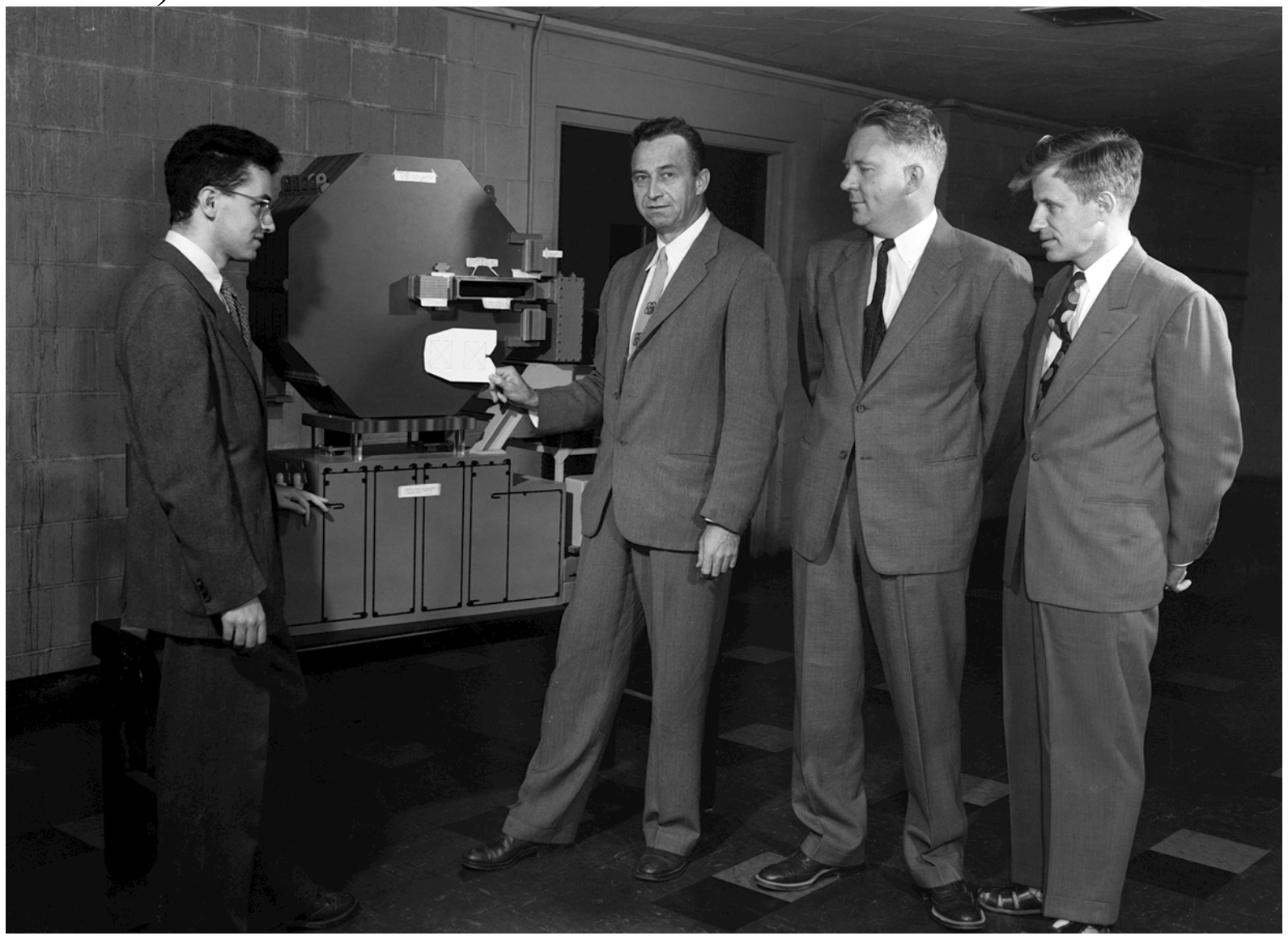

Figure 5. Courant, Livingston, Snyder and Blewett next to model of Cosmotron C-shaped magnet. Livingston holds cutout of magnet for strong focusing machine of same energy.

## The Alternating Gradient Synchrotron (AGS) Era: 1960-2002 (for HEP).

The AGS is a strong focusing synchrotron which accelerates protons to an energy of 33 GeV, or ten times that of the Cosmotron. Because of the strong focusing, the magnets of the much larger AGS (Fig. 6) contained only twice as much steel as the Cosmotron.

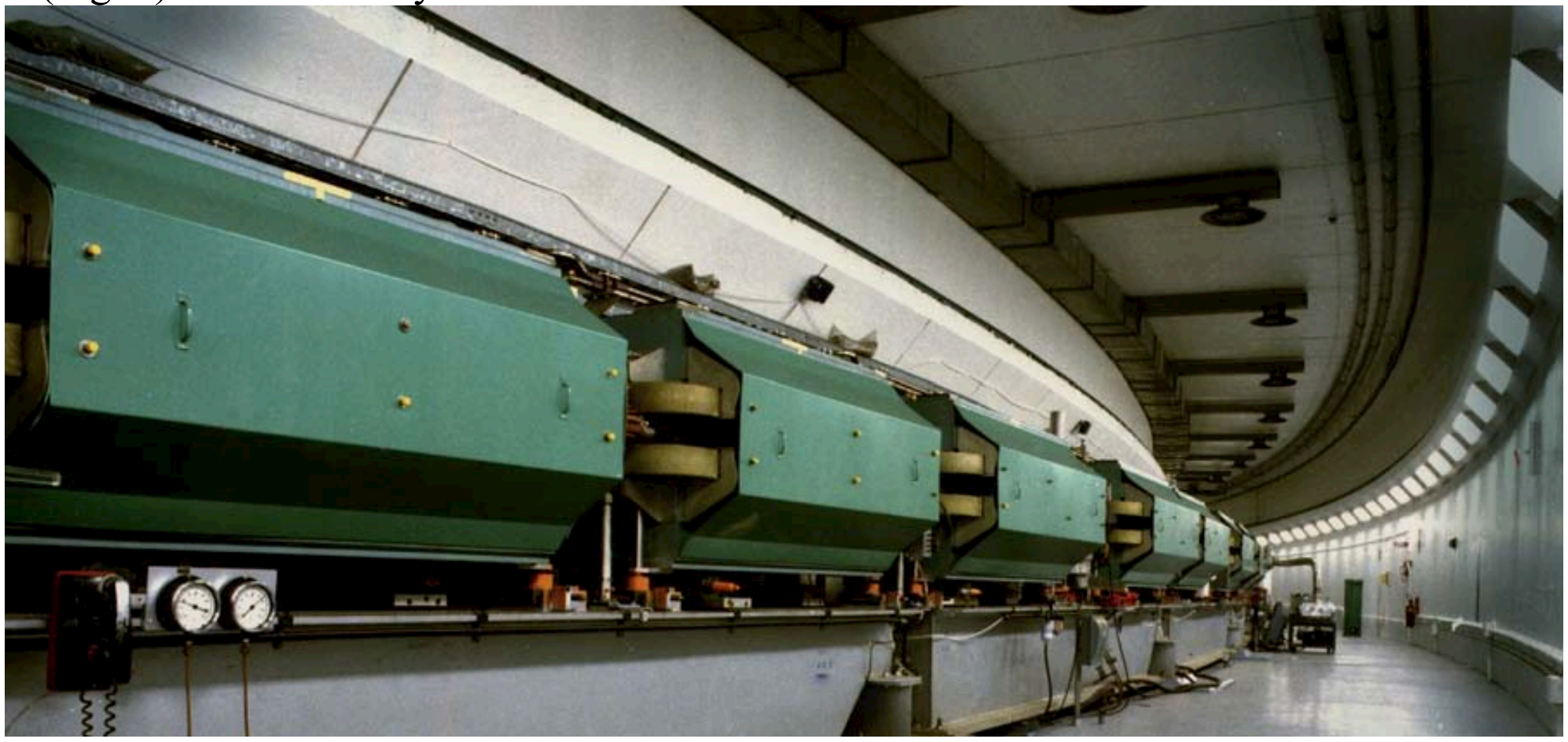

Figure 6. View of Alternating Gradient Synchrotron (AGS) machine inside its shielding enclosure. The 33 GeV machine was the highest energy accelerator in the world until 1968. The coils of the individual out-facing magnets are visible. The AGS has accelerated protons and spin-polarized protons as well as fully stripped O, Si, Cu and Au nuclei. The AGS is now used as the injector to RHIC.

The AGS has made many major discoveries in subnuclear physics, of which three received Nobel Prizes (●). Some highlights are:

- ● First neutrino beam experiment; discovery of μ-neutrino, $\nu_\mu$ (Danby 1962)
- φ-meson ($s\bar{s}$) ; $\Omega^-$ baryon ($sss$) (Barnes 1964).
- ● CP-Violation: $K^0{}_2 \rightarrow \pi + \pi - \approx 0.2$ % (Christenson 1964)
- "Drell-Yan" pairs, $p+p \rightarrow \mu + \mu - +X$
- ● J/Ψ ($c\bar{c}$) (Aubert 1974)

## Two Other BNL Neutrino Experiments-Non Accelerator.

- ν from β-decay (actually from electron capture) is left handed (Goldhaber 1958)
- ● First observation of ν from the sun (Davis 1968) Nobel Prize 2002

Figure 7a shows the entire apparatus of the experiment (Goldhaber 1958) which measured that the helicity of $\nu_e$ is left handed (i.e. its spin is opposite to its momentum) by using the reaction e-+ $Eu^{152m} \rightarrow \nu_e + Sm^{152*}$ followed by $Sm^{152*} \rightarrow Sm^{152} + \gamma$, and measuring the circular polarization of the decay γ ray by differential absorption in the magnetized iron yoke. By a clever and possibly unique choice of the isotopes used, the energy of the $\nu_e$ and γ are very close, so that the γ rays which are emitted nearly back to back from the $\nu_e$ will have the same helicity (circular polarization) as the $\nu_e$ as well as the correct energy to be 'resonantly' absorbed and re-emitted by a $Sm^{152}$ nucleus at rest, $\gamma + Sm^{152} \rightarrow Sm^{152*} \rightarrow Sm^{152} + \gamma$. The $Sm_2O_4$ resonant-scatterer is the flower pot shaped base. The re-emitted γ rays are detected in a counter under the cone-shaped Pb shield.

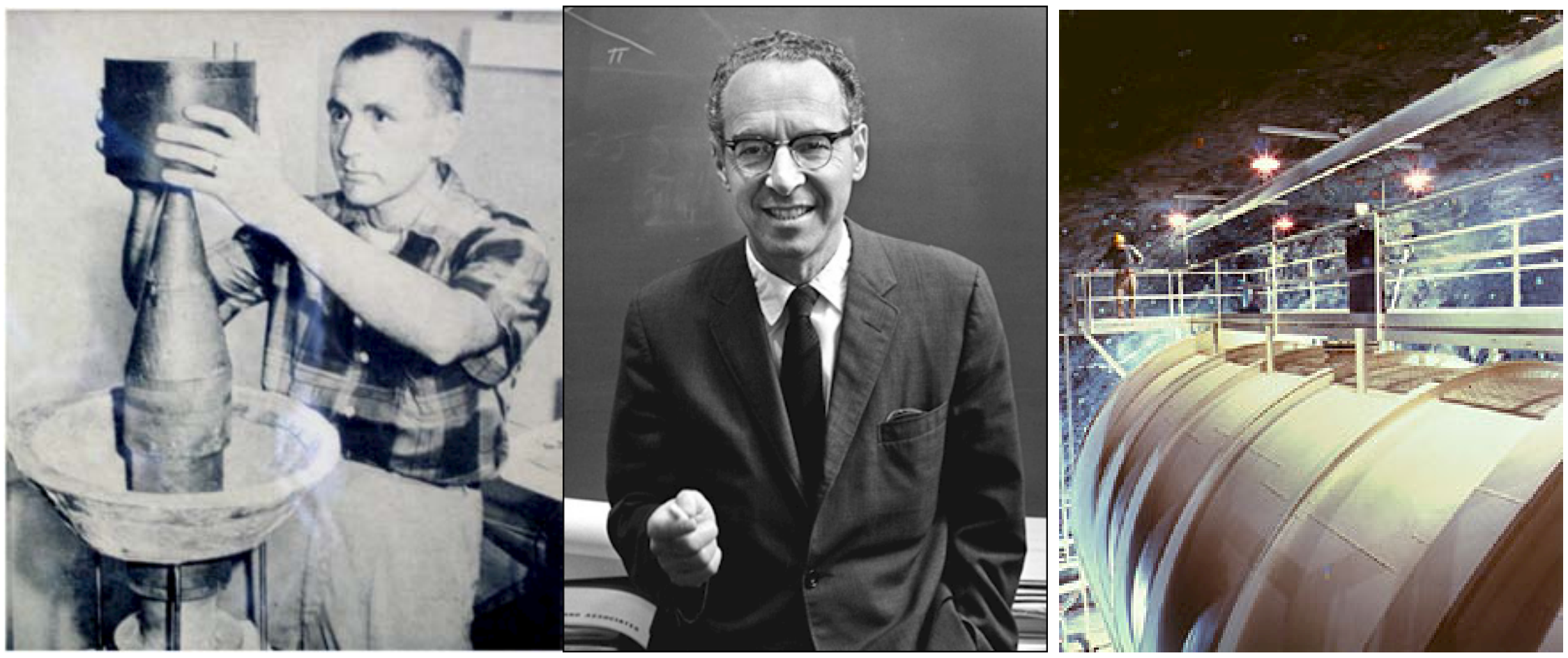

Figure 7. a) (left) Lee Grodzins c. 1957 holding the magnet of the experimental apparatus for analyzing the circular polarization of γ-rays detected by resonant scattering. The difference in counting rate due to reversing the magnetic field measures the circular polarization. The $Eu_2O_3$ source is inside the center of the magnet. The cone is a Pb absorber which restructs path of the circularly polarized γ-rays from $Sm^{152*}$ decay so that they pass through the magnetized iron on their trajectory to the basin at the bottom which is the $Sm_2O_3$ resonant-scatterer. The re-emitted γ-rays are detected in a scintillation counter under the Pb cone.
b) (center) Maurice Goldhaber c. 1967.
c) (right) Ray Davis' tank of cleaning fluid in the Homestake mine 1967. All photos courtesy BNL.

The story of Ray Davis' experiment (Figure 7c) is quicker to tell but took much longer to succeed. He put a tank of cleaning fluid (tetrachloroethylene, $C_2Cl_4$) into a deep mine from 1967-1988 and was the first to detect neutrinos from the sun via the reaction $\nu_e$ + Cl→Ar + e-. He found a deficit from the theoretical predictions. This was confirmed many years later by the 'atmospheric neutrino anomaly'. Both discoveries of a $\nu_e$ deficit shared the Nobel Prize in 2002 and were explaned by neutrino oscillations.

## The good old days!

The period discussed above, roughly from 1950—1970 were the good old days in Subnuclear Physics for several reasons. One such reason is that the discoveries were evident in photographs without resort to description of experimental details. Figure 8 shows photographs of associated production of $K^0$ and $\Lambda^0$ and the discovery of the $\Omega^-$ which appeared in the original publications. Figure 9a shows an autographed souvenir poster of the discovery of the μ-neutrino. All 8 authors and their senior technician are in the lower left panel next to a photo of their spark chamber detector. A typical event from an exposure to a neutrino beam from π-meson decay is shown above them: a long straight track which must be a muon from the reaction $\nu + A \rightarrow \mu + A'$. No events of the type $\nu + A \rightarrow e + A'$, with a short showering track, were observed. Thus the neutrino from π decay only couples to muons. This 'μ-neutrino', or $\nu_\mu$, is different from the ν from β-decay, now called $\nu_e$. Figure 9b shows Prof. Ting with a drawing of the very prominent and narrow peak, the J particle which decays to an e+e– pair. The narrow width means a long lifetime which implies a conservation law (like the strange particles), due to a new charm or $c$-quark. The J, now J/Ψ, is a bound state of $(c\bar{c})$, the hydrogen atom of QCD. These experiments in the 'good old days' were done by a dozen physicists or less, in contrast to modern experiments with ~ 400 to 4000 authors (Fig. 10).

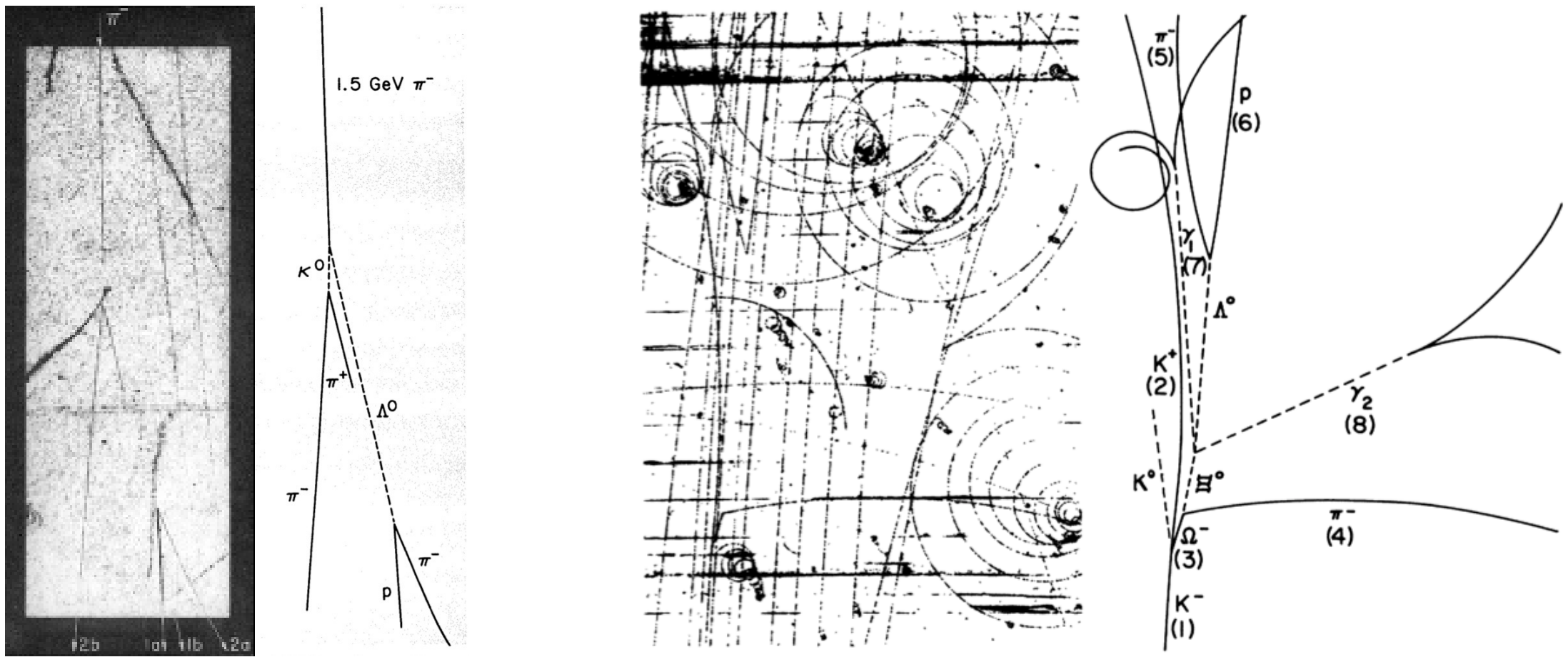

Figure 8. (left 2 panels) Hydrogen cloud chamber photograph and line diagram of event showing associated production of $K^0$ $\Lambda^0$ (Fowler 1954) with π- beam entering from the top. (right 2 panels) Hydrogen bubble chamber photograph and line diagram of event showing decay of Ω- (Barnes 1964) with K- beam entering from the bottom.

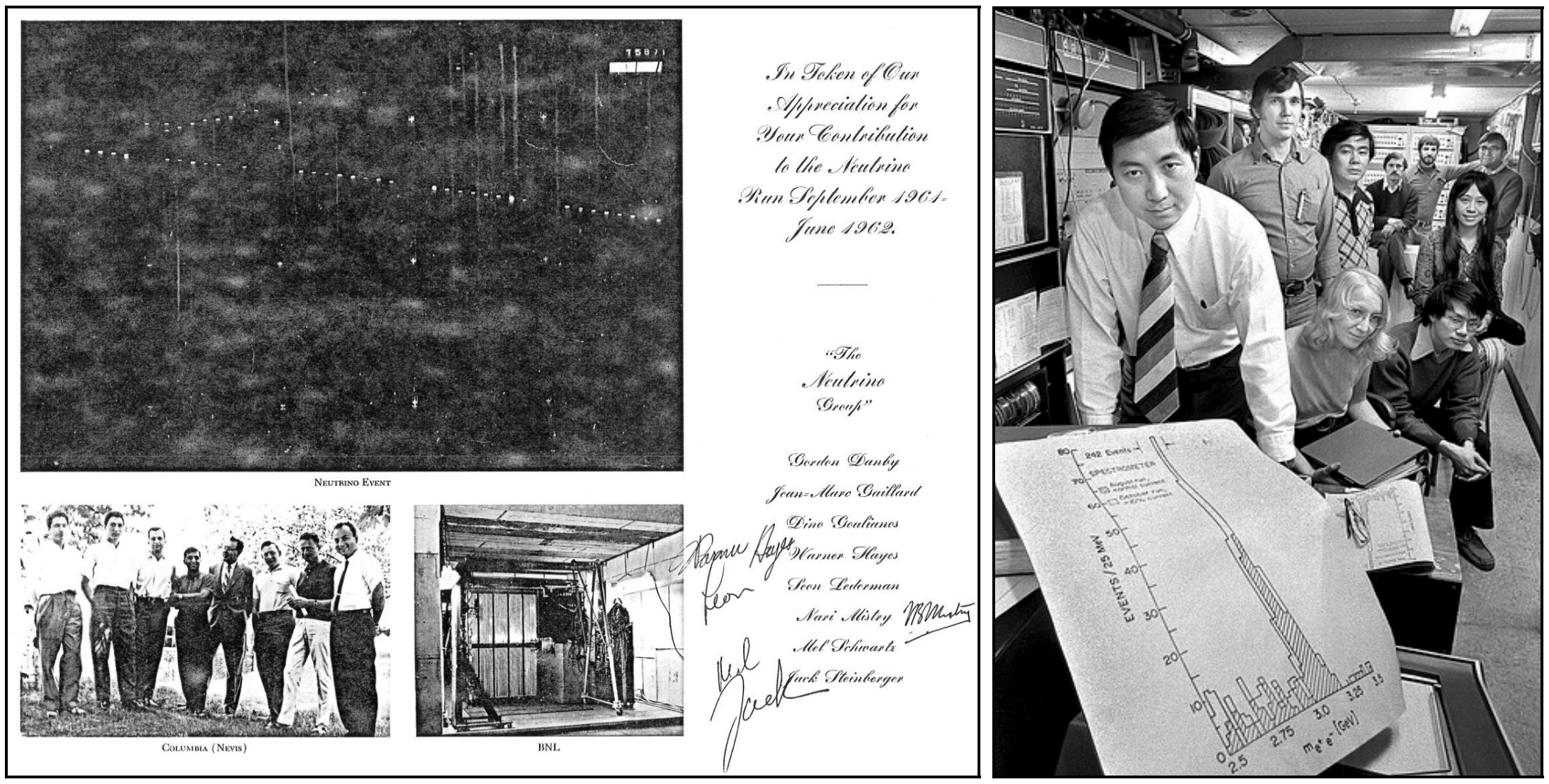

Figure 9. a) (left) Souvenir poster of the discovery of the μ-neutrino. b) (right) Prof. Ting in his counting house with members of the experiment showing a drawing of the narrow peak J which decays to an e+e– pair.

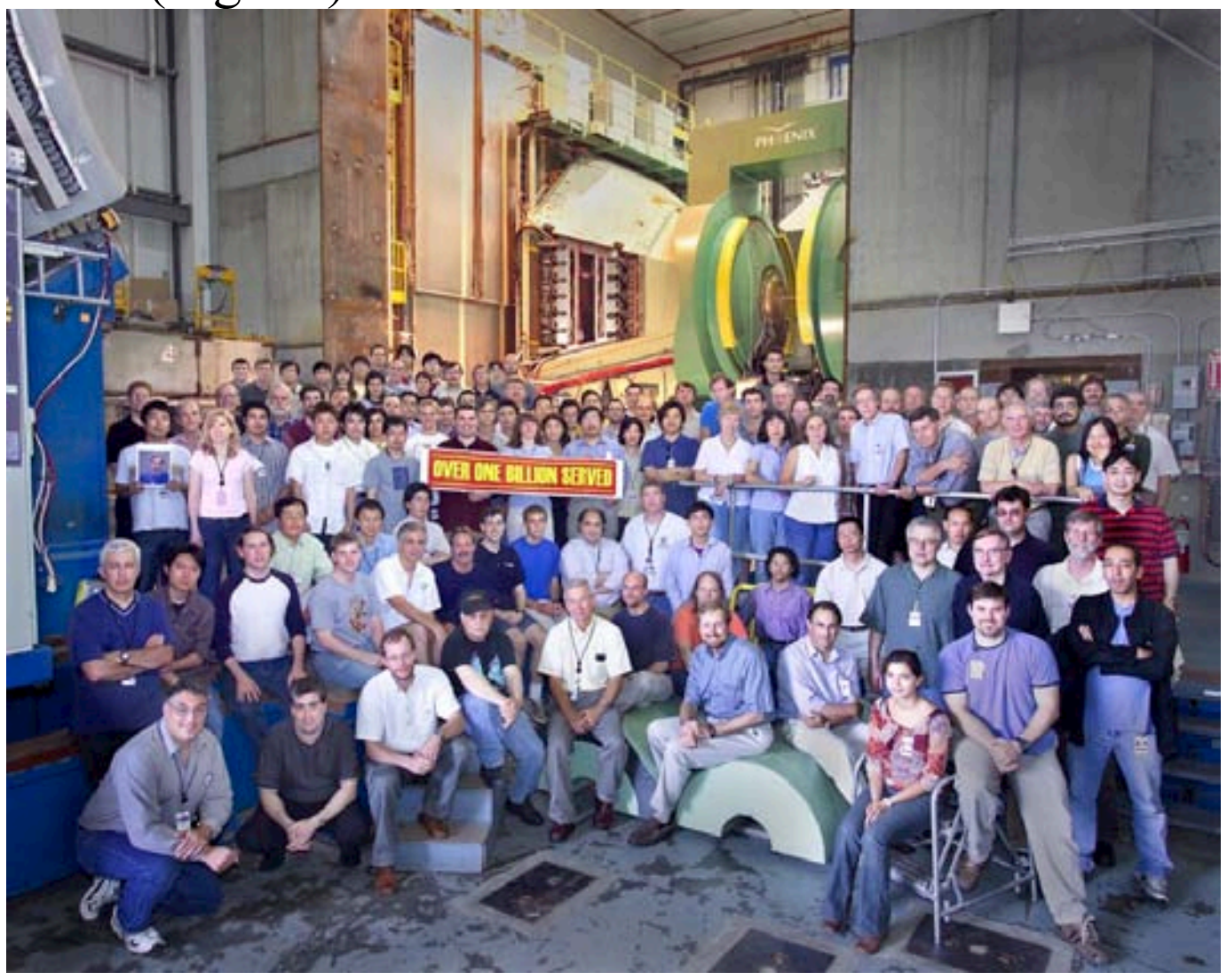

Figure 10. Some members of the PHENIX experiment at RHIC, c. 2004, in front of the open PHENIX detector.

# The Relavistic Heavy Ion Collider (RHIC)

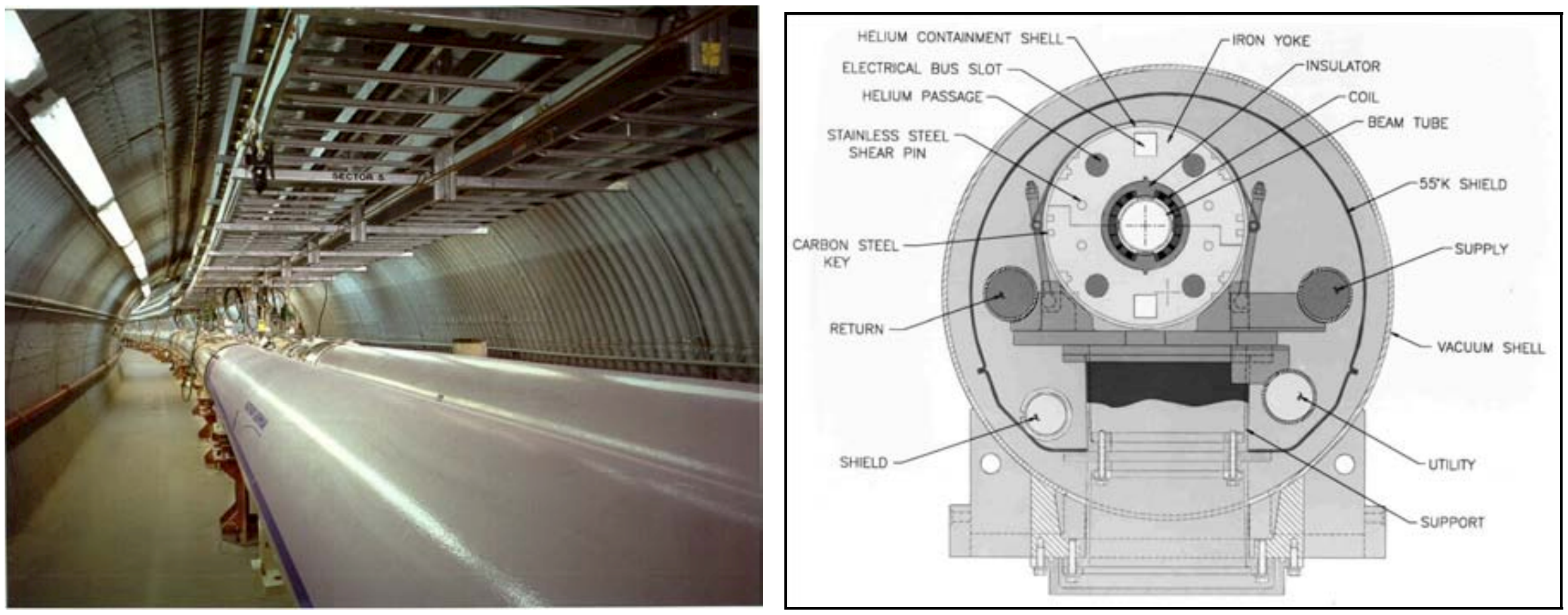

Figure 11. a)(left) Inside the RHIC enclosure—two independent rings with a total of 1740 superconducting dipole, quadrupole and corrector magnets. b)(right) Cross section of a RHIC dipole magnet viewed along the beam axis.

In the year 2000, RHIC, an accelerator-collider, with two independent rings of superconducting magnets, began operation. The Palmer magnet (Fig. 11b) used in RHIC was the basis for all post-Tevatron superconducting accelerators starting with CBA (formerly ISABELLE) including HERA, SSC, LHC: a cold iron yoke; and Cu wedges in the superconducting coils for field shaping. RHIC is a versatile accelerator which has collided Au+Au, d+Au, Cu+Cu, and polarized p-p, in runs from 2000—2011, at 12 different c.m. energies; with U+U and Cu+Au collisions scheduled for 2012.

The purpose of colliding nuclei at large nucleon-nucleon (N-N) c.m. energy ($\sqrt{s_{NN}}$) is to create nuclear matter in conditions of extreme temperature and density. At large energy or baryon density, a phase transition is expected from a nucleus as a state of nucleons containing confined quarks and gluons to a state of matter with “deconfined” (from their individual nucleons) quarks and gluons, called the quark-gluon plasma or QGP, covering a volume that is many units of the confinement length scale. The original goal of RHIC was to see whether the QGP existed and if so to measure its properties, such as phase transition temperature (T), density, equation of state, entropy, heat capacity, speed of sound, and, lately, viscosity. The experimental problems are different in A+A collisions compared to p-p because the multiplicity of soft particles is roughly A times larger in A+A than in p-p collisions, as shown with actual events in Fig. 12.

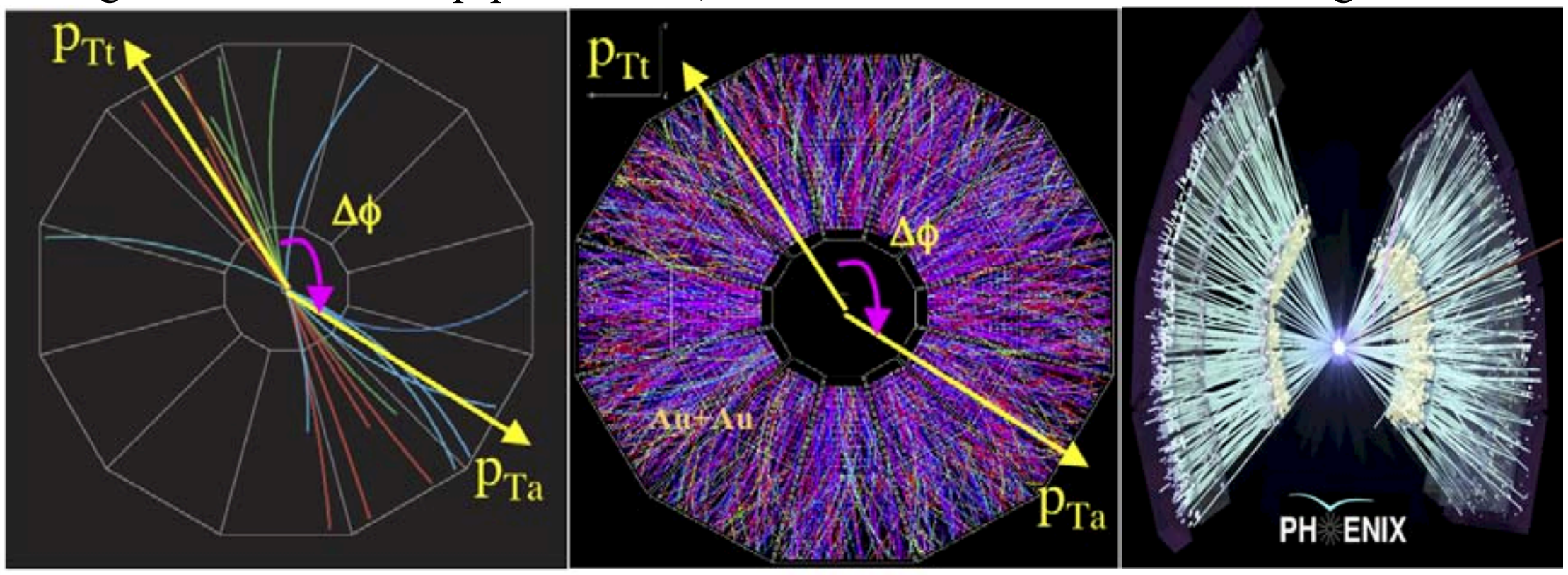

Figure 12. Collisions at nucleon-nucleon c.m. energy $\sqrt{s_{NN}}$=200 GeV a) (left) p-p collision b) (center) Au+Au central collision in the STAR detector. c) (right) Au+Au in PHENIX. Arrows $p_{Tt}$, $p_{Ta}$ at angle $\Delta\phi$ represent selected particles.

## Hard-scattering: Rutherford (1911) to the CERN-ISR to RHIC

Due the huge multiplicities in A+A collisions at RHIC, most of the analyses have been performed using single particle inclusive measurements (A+A→π +X ,where X represents all the other particles not detected) or two-particle correlations, as indicated by the arrows with transverse momentum $p_{Tt}$, $p_{Ta}$ on Fig. 12 a,b. The value of the transverse momentum, $p_T$, corresponds to the closeness or 'hardness' of the collision, with large values of $p_T$ indicating collisions at small distances, $p_T \approx 1/b$, where b is the impact parameter, or distance of closest approach of the colliding particles or nuclei (Fig 13a) . This was originally discovered by Rutherford (1911), Geiger and Marsden (1913), who scattered α-particles (He nuclei) on a gold foil and observed large angle scattering (Fig 13b) in agreement with Rutherford's calculation, a power-law (solid line) if the positive charge in an atom were all located at a central point, the nucleus, rather than spread out uniformly (dashes). In Fig 13c, the discovery of production of $\pi^0$ at large $p_T$ in p-p collisions at the CERN-ISR (CCR1973) is shown, with a power-law which depended on the c.m. energy, $\sqrt{s}$, and was qualitatively different from the exponential spectrum observed in cosmic rays for $p_T$<1 GeV/c. This represented the discovery that the point-like constituents of the proton, called partons, which are the quarks and gluons of QCD, hard-scattered strongly from each other, i.e. much larger than electromagnetically.

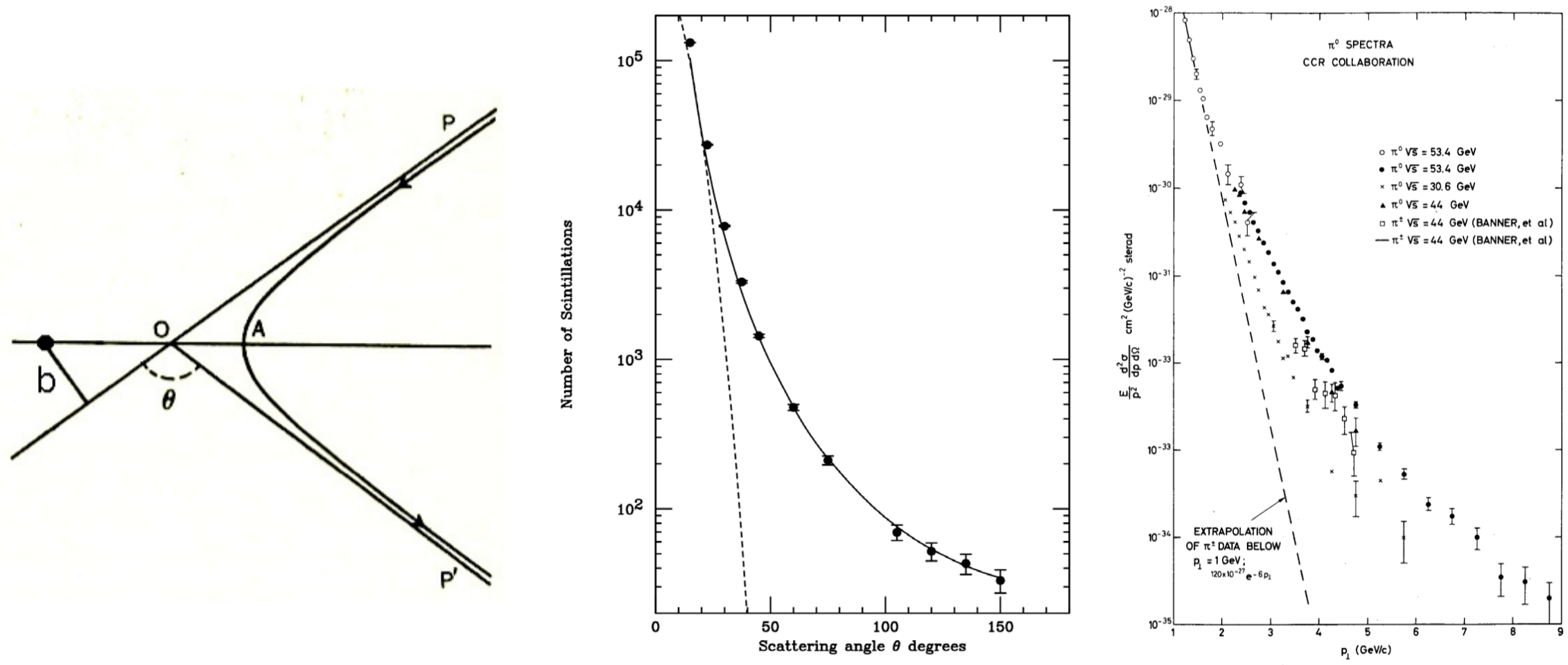

Figure 13. a) A particle P along trajectory P-O with impact parameter b from target (solid circle) scatters through angle θ )[after (Rutherford 1911)]. b) (center) Rutherford scattering cross section $1/\sin^4(\theta/2)$ (solid line) measured by Geiger and Marsden (1913) (data points); dashes indicate the expectation for a spread distribution of charge. c) (left) discovery of hard-scattering in p-p collisions at CERN-ISR (CCR 1973), $Ed^3\sigma/dp^3$ of $\pi^0$ vs. $p_T$ for several values of $\sqrt{s}$.

## Heavy Ion Collisions at RHIC—is the QGP produced?

At RHIC, in heavy ion collisions (Fig. 14a), hard-scattering of partons from the intital collision turned out to be a valuable in-situ internal probe of the medium produced. For instance, does the quark lose energy exiting the medium as sketched in Fig. 14a? If so, exactly how? etc. A baseline for for any cold-nuclear matter effects is provided by p+A (or d+A) collisions (Fig. 14b) in which no (or a very limited) medium is produced.

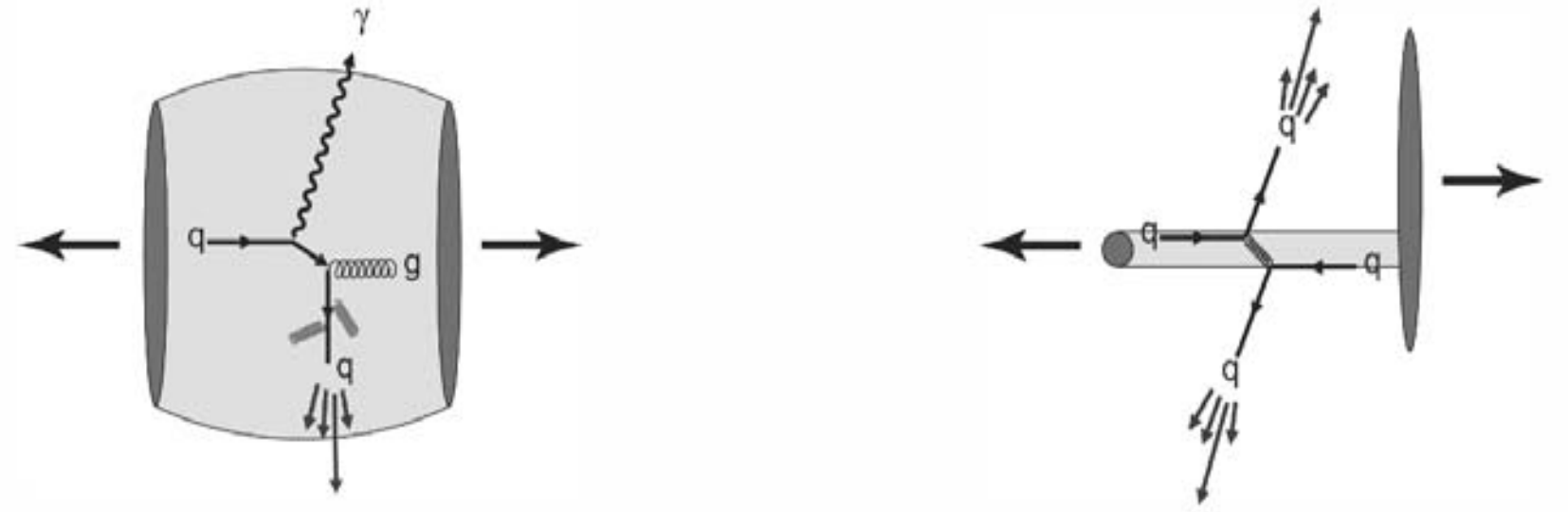

Figure 14. a) (left) Two outgoing nuclei indicated by the dark grey disks after a collision which produces a medium (light grey) in which outgoing partons from an initial hard-scattering may interact. b) (right) A p+A or d+A collision in which the medium is limited (1 nucleon wide) or non existent, so that any interaction of outgoing partons is minimal.

One of the observed distinguishing properties of the medium is that the emission of the huge number of soft particles produced is not isotropic but exhibits an asymmetry in the azimuthal distribution, $dn/dp_T d\phi=(dn/2\pi dp_T)[1+\Sigma\, 2v_n \cos n(\phi-\Phi_R)]$, represented as an expansion in Fourier harmonics, $v_n$ , where $\phi$ is the azimuthal angle in the x-y plane (Fig.15a) of a particle relative to the impact parameter vector, the x-axis ($\Phi_R$=0). This is a collective effect, known as anisotropic flow, which can not be obtained from a superposition of independent N-N collisions, but is due to the buildup of pressure in the almond shaped overlap region of the colliding nuclei. This leads to a predominant expansion of the medium in the reaction plane, which means a large $v_2$ called elliptical-flow (Fig.15b). Measurements of $v_2/n_q$ for identified hadrons, where $n_q$ is the number of constituent quarks ($n_q$=2 for mesons, $\pi$,K; $n_q$=3 for baryons p, $\Lambda$, $\Xi$), as a function of the transverse kinetic energy, $KE_T = \sqrt{p_T^2 + m^2} - m$, per constituent quark, of the particle with mass *m*, show a universal behavior (PHENIX$v_2$ 2007)[2]. The fact that flow is observed in the final state particles shows that the thermalization of the medium is rapid, so that hydrodynamic behavior can take over before the spatial anisotropy of the almond dissipates. At this early stage the hadrons are not formed so the universal behavior (Fig.15b) suggests that it is the constituent quarks rather than the particles themselves that flow. Another crucial observation (Teaney 2003) is that the persistence of flow for $p_T$>1 GeV/c, implies that the viscosity is small, perhaps as small as a quantum viscosity bound from string theory (Kovtun 2005), $\eta/s$ =1/(4$\pi$), where $\eta$ is the shear viscosity and *s* is the entropy density per unit volume. This has led to the description of the medium produced at RHIC as “the perfect fluid” (Rischke 2005), or the strongly interacting Quark Gluon Plasma (sQGP), which is a slight misnomer since the medium is a liquid, not a plasma (a gas of free quarks and gluons) as originally expected.

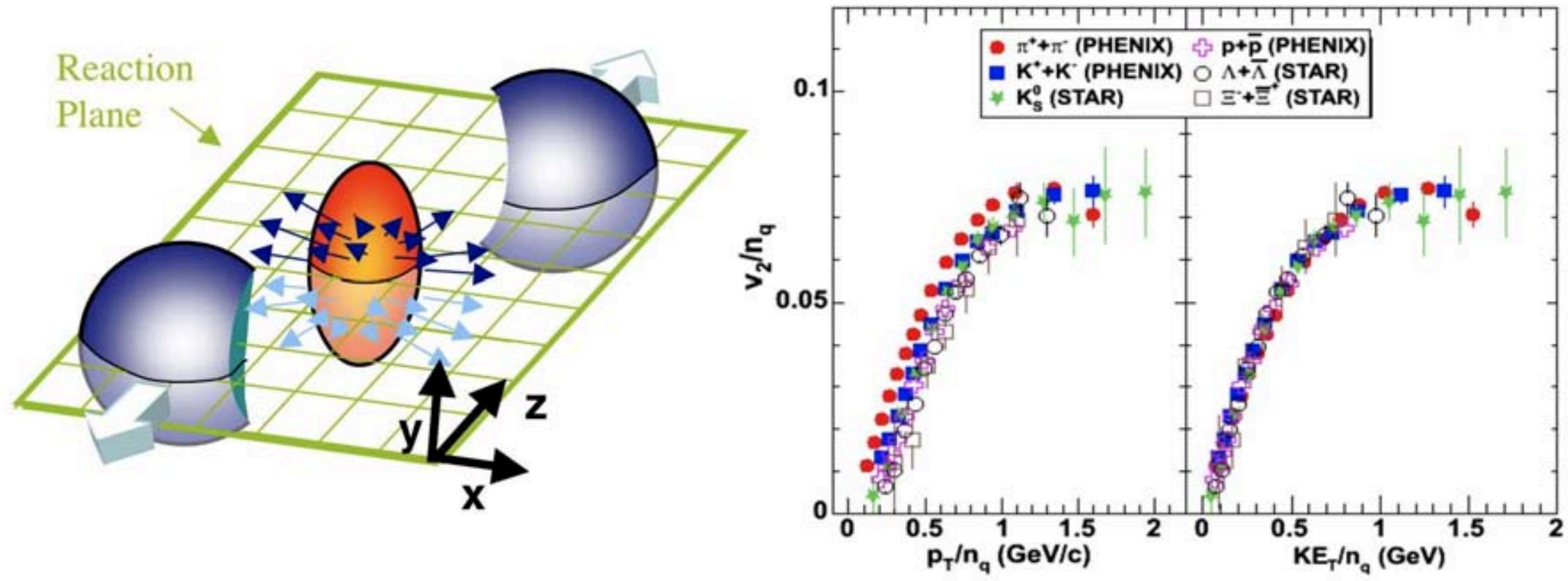

Figure 15. a) (left) Almond shaped overlap region generated just after an A+A collision where the colliding nuclei are moving along the ± z axis and the impact parameter is along the x axis. b) Measurements of elliptical flow ($v_2$) for identified hadrons plotted as $v_2$ per constituent quark vs. $p_T$, or transverse kinetic energy $KE_T$, per constituent quark.

2. See Tannenbaum (2011) for a more complete list of citations to the original publications.

## The major discovery at RHIC—Jet Quenching.

The discovery, at RHIC (Adcox 2002), that $\pi^0$'s produced at large $p_T$>3 GeV/c are suppressed in central Au+Au collisions by roughly a factor of 5 compared to point-like scaling from p-p collisions is arguably the major discovery in Relativistic Heavy Ion Physics.

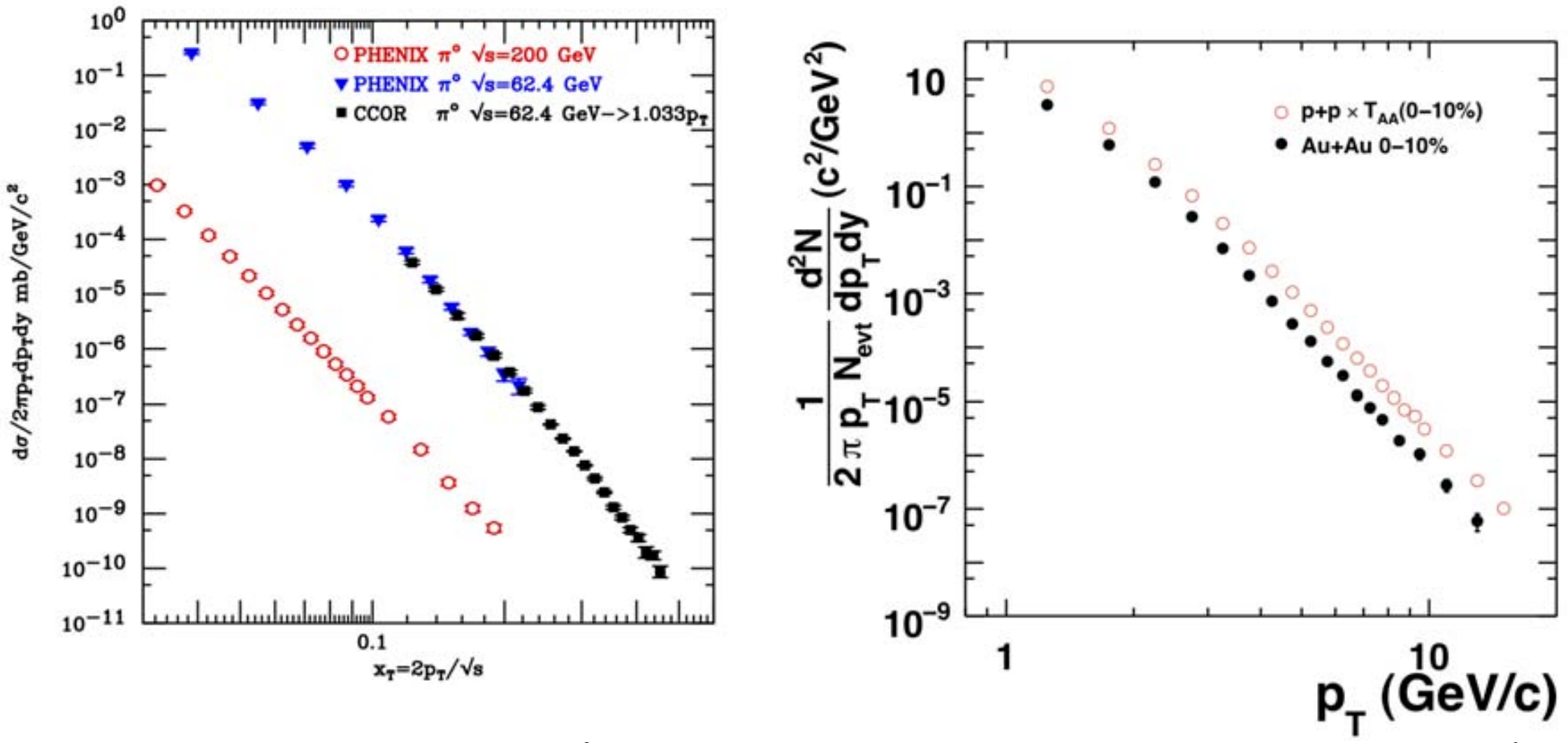

Figure 16. a) (left) Invariant cross section, $d^3\sigma/2\pi p_T dp_T dy$, where $y\approx$-ln tan($\theta$/2) is the rapidity, vs $x_T=2p_T/\sqrt{s}$ for $\pi^0$ production in p-p collisions at $\sqrt{s}$=62.4 and 200 GeV at RHIC (PHENIX 2009) compared to measurements at the CERN-ISR (CCOR 1978) at $\sqrt{s}$=62.4 GeV where the absolute $p_T$ scale has been adjusted upwards by 3.3% to agree with the RHIC data. b) (left) $\pi^0$ p-p data vs. $p_T$ at $\sqrt{s}$=200 GeV from (a), multiplied by $\langle T_{AA}\rangle$ for Au+Au central (0-10%) collisions, compared to the measured invariant yield of $\pi^0$ in Au+Au central (0-10%) collisions at $\sqrt{s_{NN}}$=200 GeV (PHENIX 2007).

In Fig.16a, the measurement of the invariant cross section, $d^3\sigma/2\pi p_T dp_T dy$, for $\pi^0$ production in p-p collisions at $\sqrt{s}$=62.4 GeV (PHENIX 2009) at RHIC is in excellent agreement with the CERN-ISR data (CCOR 1978). At $\sqrt{s}$ =200 GeV the PHENIX $\pi^0$ data follow the same (although less steep) trend as the lower energy data, with a pure power law $1/p_T^{8.10\pm0.05}$ for $p_T$ > 3 GeV/c. Since hard-scattering is point-like, with distance scale $1/p_T$<0.1 fm, much less than the size of a nucleon or nucleus, the scattering cross-section section in p+A collisions is simply A times larger than in p-p collisions and in A+A collisions, $A^2$ times larger than the p-p cross section, where A represents the number of nucleons in the nucleus. For A+A collisions in a limited range of impact parameters or "centrality" the factor is $\langle T_{AA}\rangle$, the overlap integral of the nuclear thickness functions.

In Fig.16b, the 200 GeV p-p data, multiplied by the point-like scaling factor $\langle T_{AA}\rangle$ for central (small impact parameter) Au+Au collisions are compared to the measurement of the invariant yield of $\pi^0$ in central Au+Au collisions at $\sqrt{s}$ =200 GeV (PHENIX 2007). Amazingly, the Au+Au data follow the same power-law as the p-p data but are either shifted down vertically, i.e. suppressed by a factor of ~5, or shifted lower horizontally, by ~20% in $p_T$, relative to the point-like scaled p-p data. A quantitative representation of the suppression is provided by the "nuclear modification factor" $R_{AA}(p_T)$, which is defined as the measured yield in A+A collisions at a given $p_T$, divided by $\langle T_{AA}\rangle$ times the measured cross section in p-p collisions at the same $p_T$.

In order to verify that the suppression was due to the medium produced in Au+Au collisions and not an effect in the cold matter of an individual nucleus, a measurement in d+Au collisions was performed which was so definitive that all four experiments at RHIC at the time had their results displayed on the front page of Phys. Rev. Letters (Fig. 17a).

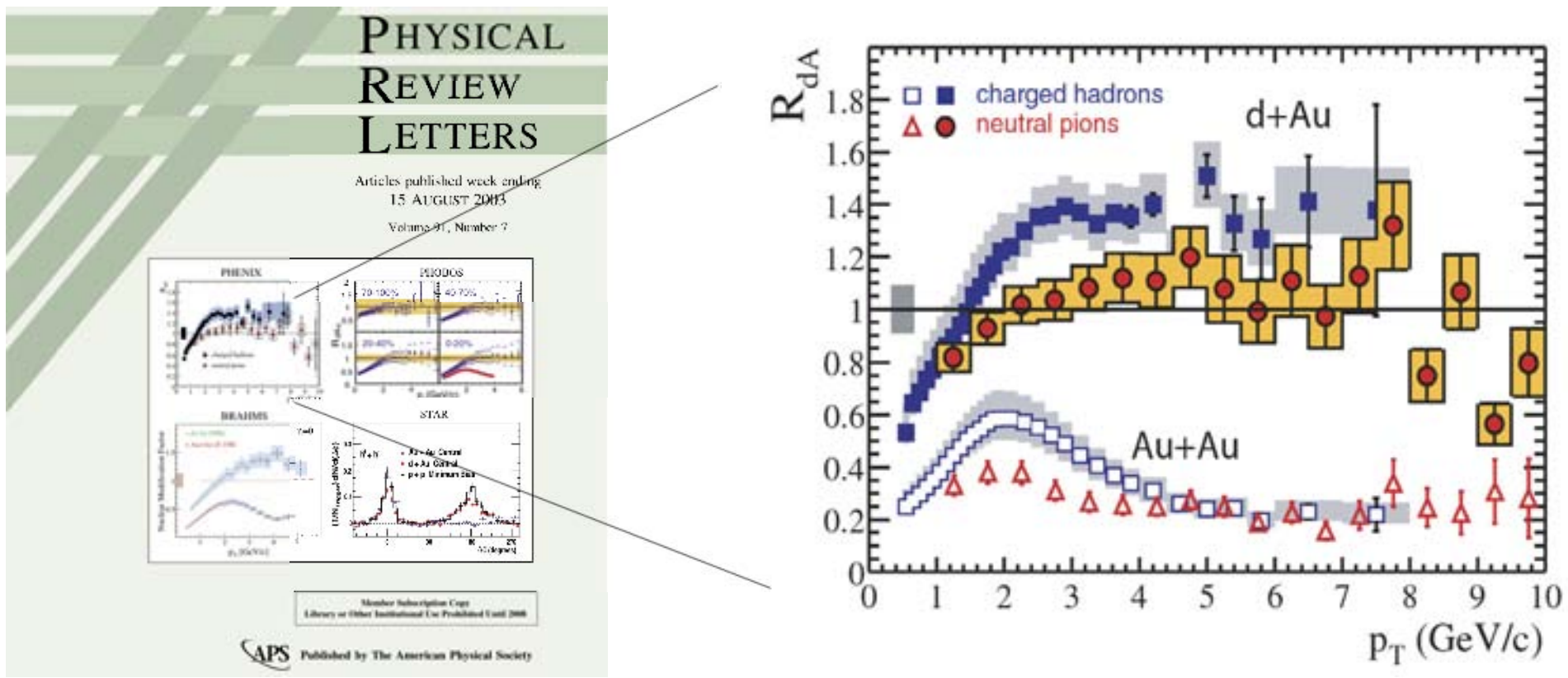

Figure 17. a) (left) Cover of Physical Review Letters of 15 August 2003 displaying the results of all 4 RHIC experiments which showed no suppression in d+Au collisions. b) (right) PHENIX results from that issue.

The conclusion from all four experiments was that there was no suppression in d+Au collisions, so that the suppression measured in A+A collisions must be due to an interaction with the medium produced, mostlikely due to radiative energy loss of the outgoing parton via gluon bremsstrahlung in the color-charged medium (Baier 2000). The PHENIX results (Fig. 17b) also showed that non-identified charged hadrons exhibited different values of $R_{AA}(p_T)$ than identified $\pi^0$ in both d+Au and Au+Au central collisions.

This illustrated the importance of systematically measuring $R_{AA}$ for identified particles.

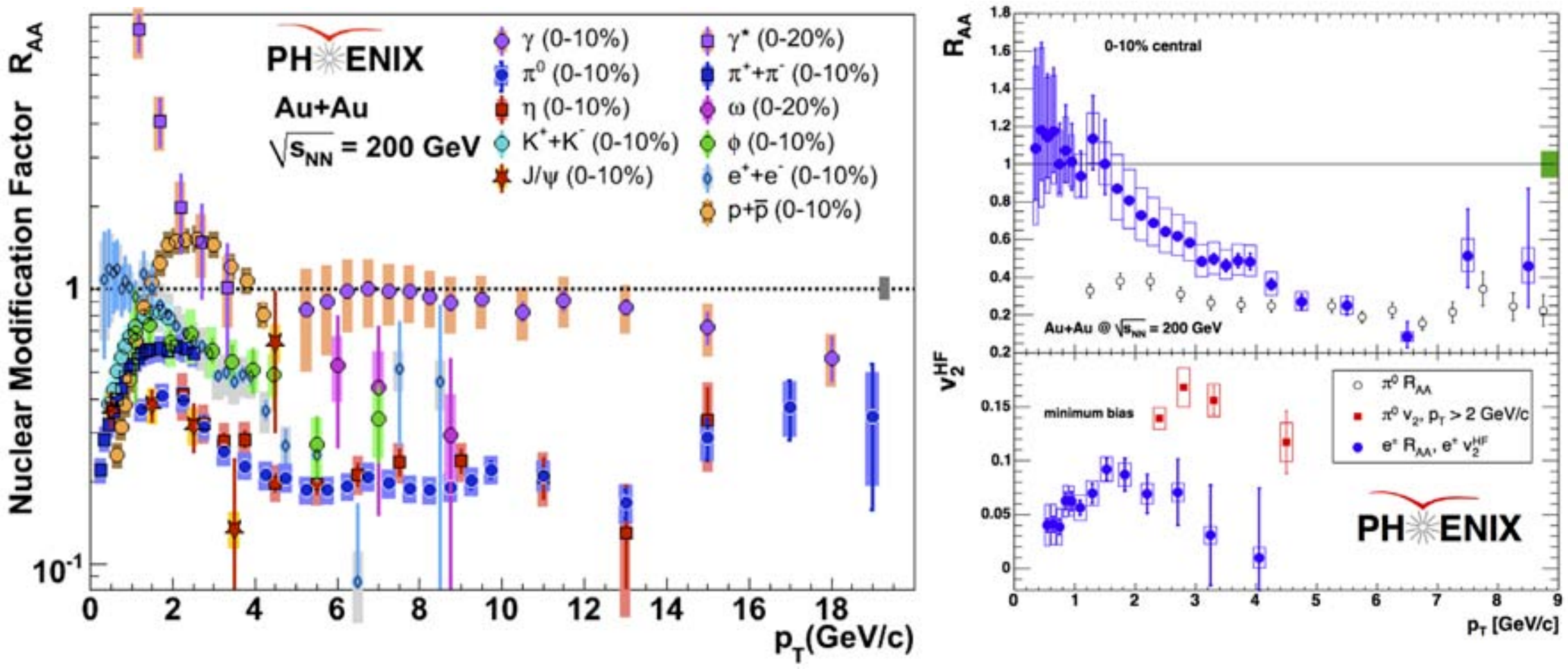

Figure 18. a) (left) PHENIX measurements of $R_{AA}(p_T)$ of the identified particles in central Au+Au collisions as indicated. b) (right) $R_{AA}(p_T)$ (top) and $v_2(p_T)$ (bottom) of single e+ or e- from b and c quark decay (PHENIXhq 2007).

PHENIX results from such systematic measurements proved to be both very interesting and exciting. Several important observations are evident from Fig.18a. 1) In the range $5 \leq p_T \leq 14$ GeV/c, γ-rays from the "inverse QCD Compton effect" g+q→γ+q (Fritzsch 1977), which do not interact with the medium since they are both electrically and color neutral, are not suppressed ($R_{AA}$=1), while $\pi^0$, which are fragments of jets from hard-scattered *u* and *d* quarks and gluons which do interact with the medium, are suppressed by a nearly constant factor of 5 ($R_{AA}\approx 0.2$), again clearly demonstrating that suppression is a medium effect on outgoing color-charged partons; 2) $\gamma^*$ (direct-$\gamma$ measured by internal conversion) for $p_T$<2 GeV/c exhibit a huge exponential enhancement over point-like scaling, shown by no other particle, which is consistent with thermal radiation from a medium with T>200 MeV (> 100,000 times hotter than the center of the sun); 3) single e+ or e- from heavy *b* and *c* quark decay are suppressed the same as $\pi^0$ ($R_{AA}\approx 0.2$) for $p_T$>4 GeV/c and exhibit flow (Fig.18b), both of which indicate a very strong interaction with the medium. This was a total surprise, a major discovery and a problem since it strongly disfavors the radiative energy loss explanation of jet-quenching because heavy quarks should radiate much less than light quarks or gluons.

One solution to this problem was offered by Professor Zichichi (2007) who proposed that the standard model Higgs Boson does not give mass to Fermions, so that "it cannot be excluded that in a QCD colored world [the sQGP] the six quarks are all nearly massless." If this were true it would certainly explain why light and heavy quarks appear to exhibit the same radiatiative energy loss in the medium. This idea can in fact be tested because the energy loss of one hard-scattered parton relative to its partner, e.g. $g + g \rightarrow b + \bar{b}$, can be measured using two particle correlations by experiments, at RHIC and LHC, in which both the outgoing *b* and $\bar{b}$ are identified by measurement of their displaced decay vertices in silicon vertex detectors. When such results are available, they can be compared to $\pi^0$-charged hadron correlations from light quark and gluon jets, for which measurement of the relative energy loss has been demonstrated at RHIC (Fig. 19). In Fig.19a, the $p_{Ta}$ spectrum of associated charged hadrons in p-p and central

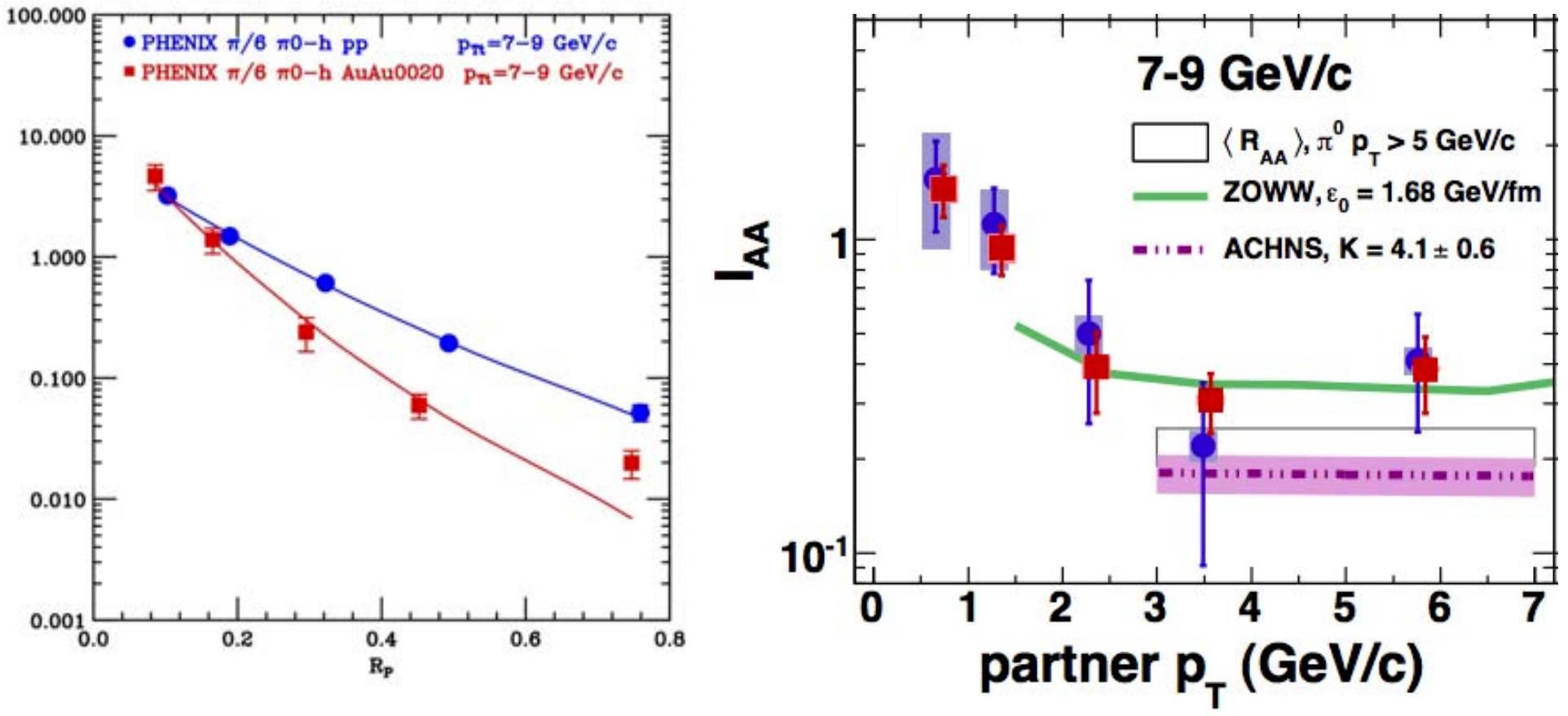

Figure 19. a) (left) $R_P$= $p_{Ta}/p_{Tt}$ distributions from p-p (circles) and central Au+Au collisions (squares) for 7<$p_{Tt}$<9 GeV/c together with fits for $R_J$. b) (right) $I_{AA}$ =ratio of Au+Au/pp data from (a) vs $p_{Ta}$ (squares) (PHENIX 2010).

Au+Au collisions is presented in the variable $R_P$= $p_{Ta}/p_{Tt}$ for trigger $\pi^0$'s with $p_{Tt}$ in the range 7—9 GeV/c. The energy loss is measured by the fact that the Au+Au spectrum is steeper than the p-p spectrum as more typically shown in the ratio of the spectra (Fig. 19b). The ratio of the away jet to trigger jet transverse momenta, $R_J$, can be found by a fit of the $R_P$ distribution of Fig. 19a to the simple formula, $dP/dR_P=N/R_J(1+R_P/R_J)^n$, where *n* is the power of the $p_T$ spectrum of the $\pi^0$ invariant yield (Fig. 16b), with result $R_J\approx 0.5$ in Fig. 19a. Zichichi's proposal could be confirmed if the same relative energy loss were also observed for *b* and *c* quarks.
Clearly, there are still exciting possibilities on the horizon in subnuclear physics.